\documentstyle[multicol,aps,prl,psfig]{revtex}

\renewcommand{\narrowtext}{\begin{multicols}{2} \global\columnwidth20.5pc}
\begin{document}

\title{Limit on Lorentz and CPT Violation of the Neutron  \linebreak
Using a Two-Species Noble-Gas Maser}
\author{D.\ Bear, R.E.\ Stoner, and R.L.\ Walsworth}
\address{Harvard-Smithsonian Center for Astrophysics, Cambridge, MA 02138}
\author{V.\ Alan Kosteleck\'y and Charles D.\ Lane}
\address{Physics Department, Indiana University, Bloomington, IN 47405}

\smallskip
\date{July 14, 2000}
\maketitle

\begin{abstract}
A search for sidereal variations in the frequency difference
between co-located ${}^{129}$Xe and ${}^{3}$He Zeeman masers
sets the most stringent limit to date on leading-order
Lorentz and CPT violation involving the neutron,
consistent with no effect at the level of $10^{-31}$ GeV.
\end{abstract}

\narrowtext
Lorentz symmetry is a fundamental feature 
of modern descriptions of nature,
including both the standard model of particle physics
and general relativity.
However,
these realistic theories are believed to be
the low-energy limit of a single fundamental theory
at the Planck scale.
Even if the underlying theory is Lorentz invariant,
spontaneous symmetry breaking might result 
in small apparent violations of Lorentz invariance 
at an observable level.
Experimental investigations of the validity of Lorentz symmetry 
therefore provide valuable tests of the framework
of modern theoretical physics.

Clock-comparison experiments 
\cite{cc1,cc2,cc3,cc4,cc5,cc6}
serve as sensitive probes 
of rotation invariance and hence of Lorentz symmetry,
essentially by bounding the frequency variation of a clock
as its orientation changes.
In practice, 
the most precise limits are
obtained by comparing the frequencies of two different co-located clocks 
as they rotate with the Earth.
Typically,
the clocks are electromagnetic signals 
emitted or absorbed on hyperfine or Zeeman transitions.

Here, 
we report on a search for sidereal variations in the frequency
of co-located ${}^{129}$Xe and ${}^{3}$He masers, 
both operating on nuclear spin-$1/2$ Zeeman transitions.
In the context of a general standard-model extension
allowing for the possibility of Lorentz and CPT violation
\cite{kp,ck},
the ${}^{129}$Xe/${}^{3}$He-maser experiment
sets the most stringent limit to date 
on leading order Lorentz and CPT violation of the neutron:  
about $10^{-31}$ GeV, 
or more than six times better than the best previous measurements
\cite{cl}.

The standard-model extension motivating this experiment 
emerges from any underlying theory that generates the standard model 
and contains spontaneous Lorentz violation 
\cite{ssb}.  
For example, 
this might occur in string theory \cite{str}.  
The standard-model extension
maintains theoretically desirable properties 
of the usual standard model
\cite{ck}.
Its formulation at the level of the known elementary particles
is a key feature enabling quantitative comparison 
of a wide array of tests of Lorentz and CPT symmetry.  
In this context,
theoretical studies have been performed 
to investigate the sensitivity of
clock-comparison experiments
\cite{cl},
tests of QED in Penning traps 
\cite{bkr}, 
experiments with a spin-polarized torsion pendulum
\cite{bk}, 
hydrogen-antihydrogen spectroscopy 
\cite{bkr2}, 
studies of photon birefringence in the vacuum
\cite{cfj,ck,jk}, 
observations of muons \cite{bkl},
measurements of neutral-meson oscillations
\cite{kp,oscill}, 
studies of the baryon asymmetry \cite{ob}, 
and observations of cosmic rays 
\cite{cg}.

In the context of the standard-model extension, 
the most sensitive prior clock-comparison experiment
is the ${}^{199}$Hg/${}^{133}$Cs comparison 
of Hunter, Lamoreaux {\it et al.}
\cite{cc6,cl}.
Recent experimental work motivated by the standard-model extension
includes Penning trap tests by Gabrielse {\it et al.} 
on the antiproton and the H$^{-}$ ion
\cite{Gab1} 
and by Dehmelt {\it et al.} on the electron and positron
\cite{Dehm1,Dehm2}.
A reanalysis by Adelberger, Gundlach, Heckel, and co-workers 
of existing data from a spin-polarized torsion pendulum experiment 
\cite{Wash1,Wash2} 
sets the most stringent bound to date on Lorentz and CPT violation 
of the electron, at about $10^{-28}$ GeV
\cite{Wash3}.  
A recent Lorentz-symmetry test using hydrogen masers 
has searched for Zeeman-frequency sidereal variations,
placing a bound on Lorentz violation at the level of $10^{-27}$ GeV 
\cite{Hmaser}.  
Together with the results of Ref.\ \cite{Wash3},
this implies an improved clean limit 
of $10^{-27}$ GeV on Lorentz-violating couplings involving the proton.
Also,
the KTeV experiment at Fermilab
and the OPAL and DELPHI collaborations at CERN have constrained
possible Lorentz- and CPT-violating effects in the $K$ and $B_d$ systems
\cite{kexpt,bexpt}.

The standard-model extension predicts 
that the leading-order Lorentz- and CPT-violating correction 
to the ${}^{3}$He-maser frequency, 
using the ${}^{129}$Xe maser as a co-magnetometer, 
is 
\cite{cl}:
\begin{equation} 
2\pi\left|\delta\nu_{J}\right| = \left|
- 3.5\tilde{b}^{n}_{J} + 0.012\tilde{d}^{n}_{J}+ 0.012\tilde{g}^{n}_{J}
\right| . 
\end{equation}
Here,
$\tilde{b}^{n}_{J}$,
$\tilde{d}^{n}_{J}$, 
and $\tilde{g}^{n}_{J}$
are small parameters characterizing the strength 
of Lorentz-violating couplings of the neutron 
to possible background tensor fields 
that may arise from spontaneous symmetry breaking 
in a fundamental theory.  
The couplings associated with 
$\tilde{b}^{n}_{J}$ and $\tilde{g}^{n}_{J}$ 
also violate CPT. 
All three parameters are linear combinations 
of more basic quantities in the underlying relativistic Lagrangian 
of the standard-model extension 
\cite{cl}.

In the analysis leading to Eq.\ (1),
the Lorentz-violating coupling 
of either the ${}^{3}$He or ${}^{129}$Xe nucleus 
has been taken as that of a single ${}^{1}S_{1/2}$ valence neutron
in a Schmidt model 
\cite{cl}.
The parameters appearing in Eq.\ (1) 
are therefore associated only with neutron couplings,
as indicated by the superscript $n$ on $\tilde{b}^{n}_{J}$,
$\tilde{d}^{n}_{J}$, $\tilde{g}^{n}_{J}$.  
Equation (1) also assumes that the applied magnetic field, 
which sets the quantization axis of the experiment, 
is directed east-west in the Earth's reference frame.
The subscript $J=X,Y$ indicates components
in the sidereal reference frame
that are orthogonal to the Earth's axis of rotation.

The design and operation of the two-species ${}^{129}$Xe/${}^{3}$He maser
has been discussed in recent publications 
\cite{3,4}. 
Here,
we give a brief review.  
The two-species maser contains dense co-located ensembles 
of ${}^{3}$He and ${}^{129}$Xe atoms.  
Each ensemble performs an active maser oscillation 
on its nuclear spin-$1/2$ Zeeman transition 
at approximately 4.9 kHz for ${}^{3}$He and 1.7 kHz for ${}^{129}$Xe 
in a static magnetic field of 1.5 gauss. 
This two-species maser operation can be maintained indefinitely.  
The population inversions for the two maser ensembles are created 
by spin-exchange collisions between the noble-gas atoms 
and optically pumped Rb vapor 
\cite{1,2}.
The ${}^{129}$Xe/${}^{3}$He maser has two glass chambers, 
one acting as the spin exchange ``pump bulb'' 
and the other serving as the ``maser bulb.'' 
This two chamber configuration permits the combination 
of physical conditions necessary for a high flux 
of spin-polarized noble gas atoms into the maser bulb, 
while also maintaining ${}^{3}$He- and ${}^{129}$Xe-maser oscillations 
with good frequency stability:
stability of about 100 nHz is typical for measurement intervals 
larger than about an hour 
\cite{4}.  
Either noble-gas species 
can serve as a precision magnetometer to stabilize 
the system's static magnetic field, 
while the other species is employed as a sensitive probe 
for Lorentz- and CPT-violating interactions.

We used the ${}^{129}$Xe/${}^{3}$He maser to search 
for a Lorentz-violation signature 
by monitoring the relative phases and Larmor frequencies 
of the co-located ${}^{3}$He and ${}^{129}$Xe masers 
as the laboratory reference frame rotated with respect to the fixed stars.  
The system was operated with the ${}^{129}$Xe maser as the co-magnetometer, 
the ${}^{3}$He maser free-running, 
and the quantization axis directed east-west in the Earth's reference frame.  
To leading order, 
the standard-model extension predicts that
the Lorentz-violating frequency shifts 
for the ${}^{3}$He and ${}^{129}$Xe maser 
are the same size and sign.
Hence,
the resultant sidereal variation of the ${}^{3}$He maser frequency
observed in the laboratory frame takes the form
\begin{equation} 
\delta\nu_{He} = \delta\nu_{X}\cos(\Omega_{s} t)
+ \delta\nu_{Y}\sin(\Omega_{s} t),
\end{equation}
where $\Omega_{s}$ is the angular frequency of the sidereal day 
\cite{sid}, 
and the parameters $\delta\nu_{J}$
given by Eq.\ (1)
represent the net effect of the Lorentz- and CPT-violating couplings 
on the ${}^{3}$He maser frequency 
with the ${}^{129}$Xe maser acting as a co-magnetometer.  
The time $t$ was measured in seconds 
from the beginning of the sidereal day 
in Cambridge, Massachusetts 
(longitude $-71.11^{\circ}$).

Data collection and analysis were performed as follows.
The ${}^{129}$Xe- and ${}^{3}$He-maser signals 
from an inductive pickup coil were buffered,
amplified,
and sent to a pair of digital lock-in detectors.  
Typical raw-signal levels were about 3 to 5 $\mu$V.  
All reference signals used in the experiment were derived
from the same hydrogen-maser clock, 
thus eliminating concerns about unmeasurable electronic phase shifts 
between the reference oscillators.  
The hydrogen maser operated on the standard hyperfine clock transition, 
and thus had no leading-order sensitivity 
to Lorentz and CPT violation 
\cite{cl,bkr2}.
Active feedback to the solenoid's magnetic field locked the phase 
of the ${}^{129}$Xe maser to that of a 1.7 kHz reference signal, 
and thereby isolated the experiment from common-mode systematic effects 
(such as stray magnetic field fluctuations)
that would otherwise shift the frequencies of the noble-gas masers
in proportion to the ratio of their magnetic moments.  
When phase locked, 
the short- and long-term frequency stability
\cite{4} 
of the ${}^{129}$Xe maser was several orders of magnitude better than 
that of the free-running ${}^{3}$He maser,
so the ${}^{129}$Xe Zeeman frequency was treated as constant 
in the data analysis.

The phase and amplitude of both maser signals were recorded 
at four-second intervals by the lock-in amplifiers
and downloaded for analysis every 23.93 hours.  
A one-sidereal day run thus contained 
approximately 21,540 evenly spaced measurements 
of the relative phases of the two masers.
The values of the two coefficients $\delta\nu_{X},\delta\nu_{Y}$ 
were computed, 
providing a measure of potential Lorentz-violation for that day's run. 
Seven additional diagnostic signals were recorded, 
including the temperatures of the pump bulb, 
maser bulb, and external resonator; 
an optical monitor of the Rb magnetization in the pump bulb; 
the broadband power emitted by the optical-pumping laser-diode array; 
the ambient room temperature; 
and the east-west component of the ambient magnetic field.  
Control loops stabilized the system temperatures to about 10 mK.
Two additional control loops were used, 
feeding back to the optical-pumping laser to reduce systematic effects 
arising from variations in the density and polarization 
of Rb in the pump bulb
\cite{Phillips}.

Small noble-gas polarization-induced frequency shifts 
were the dominant source of instability (i.e., phase drift) 
in the free-running ${}^{3}$He maser.  
For a typical one-day run, 
the linear-correlation coefficient between ${}^{3}$He phase data 
and the integrated amplitude of either maser was in the range 0.95 - 0.99.  
We admitted terms to our data-analysis model to account for this
polarization-induced phase drift.  
The effect of potential Lorentz-violating couplings on the 
evolution of the ${}^{3}$He phase
was expressed in terms of the coefficients 
$\delta\nu_{J}$ via integration of  Eq.\ (2), 
and initial reduction of each one-day run was
performed using the minimal fit model
\begin{eqnarray} 
\delta\phi_{He} &=& 
\phi_{0} + 2\pi \nu_{0} t  
\nonumber\\
&&+ 2\pi \Omega_{s}^{-1}[\delta\nu_{X}\sin(\Omega_{s} t) -
\delta\nu_{Y}\cos(\Omega_{s} t)],
\end{eqnarray}
where the coefficients $\phi_{0}$ and $\nu_{0}$ 
account for absolute phase and frequency offsets between the
${}^{3}$He maser and the ultra-stable reference oscillator. 
The reduced $\chi^{2}$ statistic for this fit model was determined, 
and then additional terms corresponding 
to quadratic and maser amplitude-correlated phase drift 
were incorporated into the model 
if they significantly improved the reduced $\chi^{2}$ 
\cite{Bev}.  
The coefficients $\delta\nu_{X}$ and $\delta\nu_{Y}$ 
for each one-day run were extracted using a linear least-squares routine 
on the best-fitting model for that day, 
which contained at most seven free parameters and thus at least
$(21500 - 7)$ degrees of freedom.  
See Fig.\ 1 for an example of the residuals from one day's data.
As a final check, 
a faux Lorentz-violating effect of known phase and amplitude was added 
to the raw data and the analysis was repeated. 
Data reduction for a given sidereal day 
was considered successful if the synthetic physics was recovered
and there was no change in the covariance matrix 
generated by the fitting routine.

\begin{figure}
\centerline{\psfig{figure=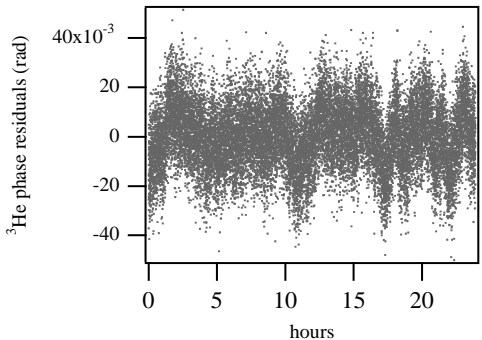,width=0.8\hsize}}
\smallskip
\caption{Typical residuals for the ${}^{3}$He phase data 
from one sidereal day,
calculated using the fit model given in Eq.\ (3).}
\label{fig1}
\end{figure}

Data for this experiment were acquired with three different maser cells
over a period of 30 days in
April 1999 (cell S3), 
24 days in September 1999 (cell E9), 
and 60 days in February-May 2000 (cell SE3, runs 1 and 2).
A total of 90 usable sidereal-day values of
$\delta\nu_{X}$, $\delta\nu_{Y}$ were obtained.
The main magnetic field of the apparatus was reversed 
about every 10 days 
to help distinguish possible Lorentz-violating effects 
from diurnal systematic variations.  
Field reversal and subsequent re-equilibration of the masers 
required approximately 24 hours.

Systematic effects resulting from possible diurnally varying 
ambient magnetic fields would not average away with field reversals.  
Thus,
the effectiveness of the ${}^{129}$Xe co-magnetometer
at eliminating such effects was carefully assessed 
before beginning data acquisition.  
Since the two maser ensembles do not have perfect spatial overlap, 
penetration of external magnetic fields through the nested magnetic shields 
and into the interaction region could induce small frequency shifts 
in the free-running ${}^{3}$He maser despite 
the presence of the ${}^{129}$Xe co-magnetometer.  
Large coils ($\sim$ 2.4 m diameter) 
surrounding the ${}^{129}$Xe/${}^{3}$He-maser apparatus 
were used to switch on and off 0.5 G external magnetic fields 
in the north-south and east-west directions.  
A bound on the ratio $|\delta\nu_{He}/\delta B_{external}|$ was obtained.  
The drifts in the ambient magnetic field near the apparatus were measured 
to be about 0.2 mG over a typical 24-hour period, 
resulting in a worst-case shift on the free-running ${}^{3}$He maser 
of less than 8 nHz, 
well below the present sensitivity of the experiment 
to Lorentz and CPT violation.
It should be noted that the relative phase between the solar and sidereal day
evolved about $2\pi$ radians over the course of the experiment 
(April 1999 to May 2000).
Hence,
diurnal systematic effects from any source would tend to
be reduced by averaging results from all measurement sets.

The measured values of $\delta\nu_{J}$ exhibited a small dependence
on the direction of the solenoidal magnetic field in the laboratory frame.
This dependence was most likely 
due to hysteresis and asymmetry in magnetic interactions
between the solenoid and the nested $\mu$-metal shields under field reversal.  
For each cell,
the data for the east and west magnetic-field orientations 
were analyzed separately to determine 
mean values and standard errors 
for $\delta\nu_{J}$,
yielding the results in Table 1.
As an example, Fig.\ 2 shows the single-day values 
of $\delta\nu_{X}$ obtained in the first run with cell SE3
in the east field orientation (SE3 E1). 

\medskip
\def\pt#1{\phantom{#1}}
\halign{\indent\qquad\quad#\hfil &\quad#\hfil &\quad#\hfil \cr
Cell &
$\delta\nu_X$ (nHz) & 
$\delta\nu_Y$ (nHz) 
\cr
\noalign{\smallskip}
S3 E& 
$\pt{-}\pt{0}95\pm 118$ & 
$\pt{-}197 \pm 114$ 
\cr
S3 W& 
$-\pt{0}43 \pm 138$ & 
$\pt{-}\pt{0}88 \pm 148$ 
\cr
E9 E& 
$-\pt{0}86 \pm 234$ & 
$-194 \pm 207$
\cr
E9 W& 
$-206 \pm 186$ & 
$-\pt{0}60 \pm 134$
\cr
SE3 E1& 
$\pt{-}100 \pm 148$ & 
$\pt{-}\pt{00}9 \pm 141$ 
\cr
SE3 W1& 
$-\pt{00}1 \pm \pt{0}88$ & 
$\pt{-}\pt{0}62 \pm 109$ 
\cr
SE3 E2& 
$-\pt{00}2 \pm 180$ & 
$\pt{-}\pt{0}68 \pm 107$ 
\cr
SE3 W2& 
$-\pt{0}35 \pm 118$ & 
$\pt{-}197 \pm 120$ 
\cr}
\smallskip
{\small
Table 1: Means and standard errors for $\delta\nu_{X}$
and $\delta\nu_{Y}$.
Results are displayed for each of the three cells (S3, E9, and SE3) 
with both east (E) and west (W) orientations of the magnetic field.  
Two runs were performed for cell SE3.}

\begin{figure}
\centerline{\psfig{figure=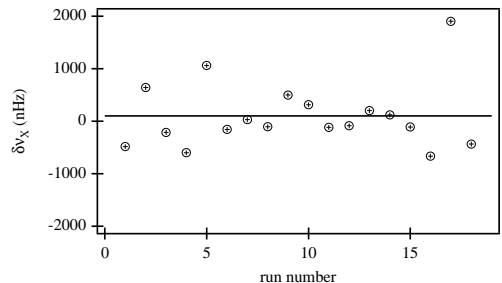,width=0.8\hsize}}
\smallskip
\caption{Values of the Lorentz-violating parameter $\delta\nu_{X}$ 
obtained with cell SE3 in the E1 orientation.
The horizontal line indicates the mean value for that data set.}
\label{fig2}
\end{figure}

The total weighted means and uncertainties 
for $\delta\nu_{X}$ and $\delta\nu_{Y}$ were then formed 
from all data sets. 
Finally, 
the results were used to extract the measured value of 
$R\equiv \sqrt{\delta\nu_{X}^{2}+\delta\nu_{Y}^{2}}$,
giving $53 \pm 45$ nHz (1-$\sigma$ level).

The size of the coefficients in Eq.\ (1)
indicates that the ${}^{129}$Xe/${}^{3}$He-maser experiment 
is most sensitive to the Lorentz- and CPT-violating coupling 
associated with $\tilde{b^{n}_{J}}$.
Under the assumption of negligible 
$\tilde{d}^{n}_{J}$ and $\tilde{g}^{n}_{J}$
\cite{cl},
the above experimental result for $R$ 
corresponds to a value for 
$\tilde{b}^n_\perp\equiv\sqrt{(\tilde{b}^n_X)^2+(\tilde{b}^n_Y)^2}
=(4.0\pm 3.3) \times 10^{-31}$ GeV.
This result is consistent with no Lorentz- and CPT-violating effect,
given reasonable assumptions about the probability distribution
for $R$
\cite{R_distribution}.
It represents the most stringent limit to date on possible 
Lorentz- and CPT-violating couplings involving the neutron
and is more than six times better than the best previous measurements 
\cite{cl}.  

We are planning improved Lorentz and CPT tests using noble-gas masers.  
Upgrading laser and temperature control and
acquiring a larger data set could better 
the present Lorentz and CPT constraint
from the ${}^{129}$Xe/${}^{3}$He system
by up to an order of magnitude.  
Also, 
a new two-species Zeeman maser 
using ${}^{3}$He and ${}^{21}$Ne might provide
even greater improvements to constraints on neutron parameters
\cite{cl,Stoner}.  

David Phillips, Mark Rosenberry, 
Federico Can\'e, Timothy Chupp, Robert Vessot 
and Edward Mattison helped greatly with this project.  
Development of the ${}^{129}$Xe/${}^{3}$He Zeeman maser was
supported by a NIST Precision Measurement Grant.  
Support for the Lorentz-violation test was provided by 
NASA grant NAG8-1434, 
ONR grant N00014-99-1-0501,
and the Smithsonian Institution Scholarly Studies Program.
This work was supported in part by
the DOE under grant DE-FG02-91ER40661.

\bibliographystyle{prsty}

\begin{references}
\bibitem{sid} 
One sidereal day is approximately one solar day
scaled by $(1 - 1/364.24)$,
or about 23.93 hours.
\bibitem{R_distribution} 
The probability distribution for 
$R$
is formed by multiplying the distributions 
for $\delta\nu_{X}$ and $\delta\nu_{Y}$, 
converting to polar coordinates, 
and integrating over the polar angle.  
In the case where $\delta\nu_{X}$ and $\delta\nu_{Y}$ 
have zero mean value and the same variance $\sigma$, 
the probablity distribution takes the form 
$P(R) = \sigma^{-2}R \exp(-r^{2}/2\sigma^{2})$,
with the most probable value of R occurring at $R = \sigma$.
\end{references}

\end{multicols}
\end{document}